\documentclass[global,twocolumn]{svjour}
\usepackage{epsfig}
\usepackage{graphics}
\usepackage{epsfig}
\usepackage{times}
\usepackage{amsmath}
\usepackage{mathtools}
\usepackage[english]{babel}
\begin{document}
\DeclareGraphicsExtensions{.pdf,.png,.jpg,.eps,.tiff}

\title{Optimum electrode configurations for fast ion separation in microfabricated surface ion traps}
\author{Altaf H. Nizamani  and
Winfried K. Hensinger }

\institute{Department of Physics and Astronomy, University of Sussex, \\
Falmer, East Sussex, Brighton, BN1 9QH, UK}

\mail{Winfried Hensinger\\
e-mail: W.K.Hensinger@sussex.ac.uk}

\date{\today}
\maketitle

\begin{abstract}

For many quantum information implementations with trapped ions, effective shuttling operations are important. Here we discuss the efficient separation and recombination of ions in surface ion trap geometries. The maximum speed of separation and recombination of trapped ions for adiabatic shuttling operations depends on the secular frequencies the trapped ion experiences in the process. Higher secular frequencies during the transportation processes can be achieved by optimising trap geometries. We show how two different arrangements of segmented static potential electrodes in surface ion traps can be optimised for fast ion separation or recombination processes. We also solve the equations of motion for the ion dynamics during the separation process and illustrate important considerations that need to be taken into account to make the process adiabatic.

\end{abstract}

\maketitle

\setcounter{secnumdepth}{1}
\section{Introduction}


Significant progress has been made in quantum information processing with trapped ions \cite{Cirac,Monroe2,Wineland2,Haeffner}, including entanglement gates \cite{Monroe,Kaler},  teleportation \cite{Bennett,Barrett,Riebe} and quantum simulation \cite{Leibfried,Porras,Johanning}.

However it is not easily feasible to manipulate many qubits in a single trapping region. It would be useful if qubits can be stored in separate trapping regions (\emph{memory zones}) and only be brought together in a single trap (\emph{processor zone}) when quantum operations are required \cite{Wineland,Cirac2,Kielpinski,Steane2}. Shuttling within an array of ion traps has been demonstrated successfully in linear arrays and through junctions \cite{Hensinger,Schulz,Huber,Blakestad,Amini}. Two ions have also been reordered by rotation within a linear trap section \cite{Splatt}. Separation of two pairs of ions having different masses was also demonstrated \cite{Jost}. Ions are separated, re-combined and transported across the different zones of an ion trap array by means of time-varying potentials on control electrodes. How to optimise electrode geometries for efficient ion separation and recombination has been discussed by Home and Steane \cite{Home} in general. House \cite{House} studied surface-electrode ion traps analytically. Hucul \emph{et al.} \cite{Hucul} and Reichle \emph{et al.} \cite{Reichle} have discussed the energy gain of trapped ions due to ion transport. We extend their findings to the ion separation process in surface trap arrays, providing a detailed description how such arrays need to be optimised to allow for efficient separation. Furthermore, we analyse the dynamics of the separation process by solving the equations of motion and present a description about the considerations that need to be taken in order to make the process adiabatic.\\
There are two types of trap geometries, asymmetric ion traps \cite{3Pearson,Seidelin,Britton,Allcock}, where the electrodes lie in a plane and the ions are trapped above that plane and symmetric ion traps \cite{Rowe,Blatt,Stick,Hensinger,Blakestad}, where the electrodes are symmetrically positioned around the position of the trapped ion. It is important to scale these architectures to trap and shuttle hundreds of ions for any useful quantum information processing to occur. This article focusses on optimal geometries for asymmetric ion traps. Modern microfabrication techniques are a promising approach to build such scalable ion trap arrays in which ions will be brought together and separated many times in processor zones to perform the gate operations. We will show that this is best attainable when the trap features are designed at the scale of the ion-electrode distance. In this article we discuss the optimisation for the particular case of surface ion trap arrays. The speed of the adiabatic shuttling operation can be enhanced by maximising the secular frequency during separation and recombination inside the trap array \cite{Hucul}. The secular frequency depends on the applied voltages on the electrodes, the geometry of the traps and it typically increases for smaller ion-electrode distance. However, at smaller scales, motional heating of ions becomes significant due to anomalous heating \cite{Turchette,Deslauriers,Jaroslaw}. Cryogenic operation of ion trap chips may allow for small ion-electrode spacings as it is known to significantly suppress anomalous motional heating \cite{Deslauriers,Jaroslaw}.\\
In surface ion traps the ion-electrode distance (ion height) depends on the size and configuration of the electrodes \cite{3Pearson,House}. The average ion life-time in a trap depends on the trap depth. One of the main challenges in surface traps is to achieve higher trap depths at larger ion-electrode distance, since such traps typically offer depths of about 1$\%$ that of multi layer symmetric traps of comparable dimensions \cite{2Chiaverini} and the magnitude of the voltage that can be applied is limited by the actual fabrication process.\\
In this article we discuss how to design a surface trap array in which ions can be trapped at a maximum trap depth for a given ion-electrode distance and can be brought together and separated rapidly by adjusting static voltages on electrodes while maintaining the highest possible secular frequencies. The trap depth and secular frequencies are dependent on the applied voltages and geometric factors of surface trap geometries. As the applied voltages are limited by power dissipation and breakdown voltage of the trap electrodes, it is important to optimise the trap depth and secular frequencies by adjusting the dimensions of the electrodes.
In Sec. \ref{sec:TD} we show that the trap depth may be optimised at a given ion height by adjusting the size and configuration of the electrodes. In Sec. \ref{sec:beta}, we show how to maximise the secular frequency during the separation and recombination shuttling processes by adjusting the widths of the static potential electrodes for ion transportation in general, and fast ion separation processes in particular. In Sec. \ref{sec:dynamics}, we discuss the dynamics of the separation process. We discuss constraints in the design of realistic trap arrays. We then compare two fundamental designs and present a guide to accomplish fast and adiabatic ion separation.


\section{Optimisation of trap depth}
\label{sec:TD}

\begin{figure}[!tb]
\includegraphics [width=85mm] {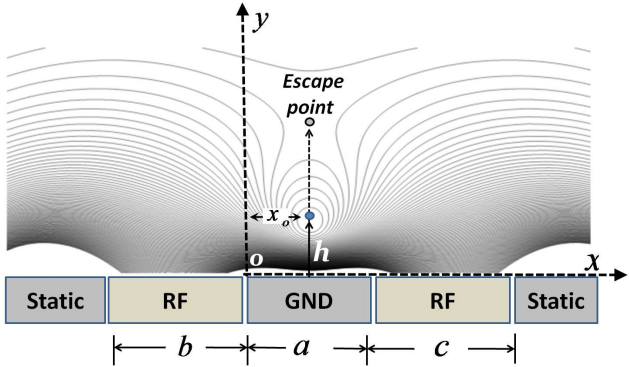}
\caption{Illustration of the pseudopotential created above the surface of the trap electrodes when an rf voltage is applied on the rf electrodes while keeping the other electrodes at rf ground. $x_o$ is the horizontal position of the ion with respect to the central ground  electrode, \emph {h} is the height of a trapped ion above the central ground electrode, and the escape point (turning point) for the ion is located beyond \emph {h}. \emph{b} and \emph{c} are the widths of the rf electrodes and \emph{a} is the separation between the rf electrodes. }
\label{contour}
\end{figure}

In a typical three-dimensional rf Paul trap the rf field provides trapping in the \emph{x} and \emph{y} dimensions (transverse axes) and a static potential provides confinement in the \emph{z}-direction. The effective potential in all three directions is given by \cite{Dehmelt,Hucul},
\begin{equation}\label{eq:iondynamics}
\Psi(\chi,t)=\frac{e^2 V^2_{rf}}{4m\Omega^2_{rf}}|\nabla\Theta_{rf}(\chi)|^2+e \sum_i V_i(t)\Theta_i(\chi)
\end{equation}
where, \emph{e} and \emph{m} are the charge and mass of the ion being trapped, $ \Theta_{rf}(\chi)$ is the instantaneous electric rf potential when $V_{rf}=1$ V and $ \Theta_i(\chi)$ is the static electric potential produced by the $i$th static potential electrode when $V_i=1$ V and $\chi$ is the position vector. $V_{rf}$ is the peak rf voltage applied on the rf electrodes with drive frequency $\Omega_{rf}$ and the coefficient $V_i(t)$ is the time varying voltage applied on the \emph{i}th control electrode. The first part of Eq. \ref{eq:iondynamics} represents a pseudopotential which can be created in a trap by applying an rf voltage on the rf electrodes while keeping the other electrodes at rf ground. In a surface trap, the position of the minimum of the pseudopotential or \emph{rf node} where the ions can be trapped, is located at a distance \emph{h} (ion height) in the \emph{y}-direction and $x_o$ in the \emph{x}-direction, as shown in Fig. \ref{contour}. The escape point or turning point of the pseudopotential shown in Fig. \ref{contour} is located above the trapping position. The position of the ion and the turning point can be found by calculating where the gradient of the pseudopotential is zero. In absence of any static potential, the trap depth defined as the amount of energy needed for an ion to escape, can be represented by the difference between the pseudopotential at the \emph{rf node} (typically zero) and the turning point. In Fig. \ref{contour}, the trap electrode dimensions are labelled as widths of the rf electrodes \emph{b} and \emph{c} and separation between the rf electrodes \emph{a}.\\
Small gaps between the trap electrodes in realistic geometries usually have negligible effects on trap parameters \cite{Roman,House}. Therefore, the basis functions for the trap electrodes can be calculated using the analytical model incorporating the gapless plane approximation \cite{House}. As House \cite{House} suggested,  if the origin of the coordinate system is located between the ground and left rf electrode as shown in Fig. \ref{contour} and the segmented static potential electrodes are considered infinitely long (in the $x$-direction), the \emph{rf node} can be found to be positioned at, $x_{o}=ac/(b+c)$ and $h=\sqrt{abc(a+b+c)}/(b+c)$ \cite{House}. The ion should be at a reasonably large distance from the trap electrodes to reduce the effect of anomalous heating of the ion \cite{Jaroslaw,Deslauriers} and provide good laser access. The later requirement may be alleviated via the introduction of slots in the substrate that allow for optical access \cite{Leibrandt,VanDevender}. First we show how to maximise the trap depth for a given ion height $h$.\\
Building on the discussion given by House \cite{House}, we re-express the trap depth $\Xi$ in terms of given ion height $h$ and the geometric factor $\kappa$ for a given ion mass $m$ and rf voltage $V_{rf}$,
 \begin{equation}
 \Xi=\frac{e^{2} V^{2}_{rf}}{\pi^2 m \Omega_{rf}^{2}h^2}\kappa
\label{TD}
\end{equation}
where $\kappa$ is described as
\begin{equation}
\kappa=\bigg[\frac{2\sqrt{a b c (a+b+c)}}{(2a+b+c)(2a+b+c+2\sqrt{a(a+b+c})}\bigg]^2.
\label{kappa}
\end{equation}

\begin{figure}[!t]
\center
\includegraphics [width=85mm] {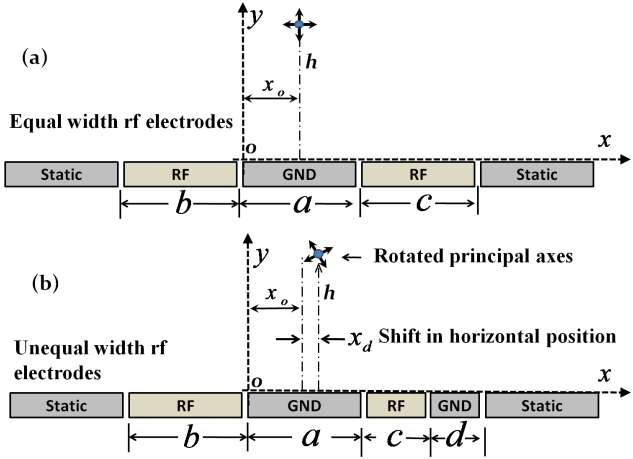}
\caption{Illustration of the principle axis rotation. (a) Orientation of the principle axis when equal width rf electrodes ($b=c$) are used. (b) The principal axis is rotated by using unequal width rf electrodes ($b\neq c$). The horizontal shift $x_d$ in the trapping position is caused by the configuration of the rf electrodes. To keep the static potential electrodes symmetric around the trapping position in the axial direction, an extra ground electrode of width $d$ is inserted between the narrower rf electrode and the static potential electrodes.}
\label{Fig:rotation}
\end{figure}

Eq. (\ref{TD}) shows that the trap depth for a given ion height $h$, can be maximised by optimising the geometric factor $\kappa$ which is defined in Eq. (\ref{kappa}).
By choosing the appropriate electrode widths, the trap geometries can be optimised to achieve the maximum trap depth for a given ion height.
Efficient laser cooling of an ion along all three directions of motion can only be achieved if the laser wave vector $k$ has a vector component along all three principal axes. Therefore, in most ion trap experiments, rotation of the principal axes is required for effective laser cooling of an ion in all three directions \cite{Itano}. One of the techniques to achieve rotation of the principal axes is to use asymmetric rf electrodes which have different widths \cite{Wesenberg} as shown in Fig. \ref{Fig:rotation}.
Unequal widths $(b\neq c)$ of the rf electrodes cause planar asymmetry in the $x$-axis and set off the nonuniformity in the confinement field when equal static voltages are applied on opposite static potential electrodes. This issue can be managed by introducing an extra ground electrode of width \emph{d} between the narrower rf electrode and the segmented electrodes parallel to the rf electrode as shown in Fig. \ref{Fig:rotation}. The width $d$ of the ground electrode (in the gapless approximation) may be chosen as the difference in the widths of the rf electrodes ($\Delta w= |b-c|$) plus the shift in the horizontal position $x_d$ of the \emph{rf node}, caused by the unequal widths of the rf electrodes. The width $d$ can then be calculated as
\begin{equation}
d=\Delta w+x_d =\Delta w+ \big(\frac{ac}{b+c}-\frac{a}{2}\big).
\end{equation}

\begin{figure}[!b]
\includegraphics [width=85mm] {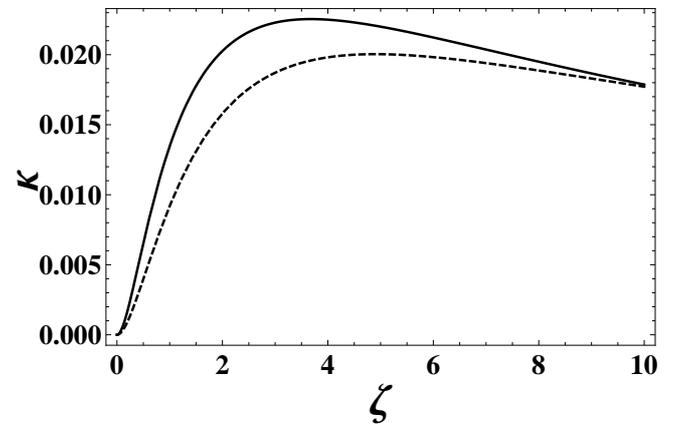}
\caption{For a given ion height, the trap depth geometric factor $\kappa$ can be maximised for equal width rf electrodes ($b=c$) when the ratio between  rf width and separation $\zeta \approx 3.68$ (solid curve). In the case when the rf electrodes widths are unequal and ($c=b/2$), $\kappa$ is maximised when the ratio is $\zeta \approx 4.9$ (dashed curve).}
\label{Fig:ratio}
\end{figure}

House \cite{House} optimised trap depth for a given rf electrode separation $a$. However, we believe it is more useful to derive values for a given ion height since ion height is a major constraint for many experiments due the occurrence of anomalous heating. The ratio $\zeta =b/a$ between the rf electrodes widths, \emph{b} and the separation between rf electrodes, \emph{a} is useful to characterise $\kappa$ at a given ion height $h$. Using the ratio $\zeta$, the geometric factor $\kappa$ can be parameterised for equal and unequal width rf electrodes as following
\begin{equation}
 \kappa = \begin{dcases*}
        \frac{\zeta^2 (1+2 \zeta)}{4(1+\zeta)^2(1+\zeta+\sqrt{1+2\zeta})^2}  & when $c=b$ \\ \\
        \frac{4 \zeta^2 (2+3\zeta)}{(2+1.5\zeta)^2(4+3\zeta+4\sqrt{1+1.5\zeta})^2} & when $c=b/2$
        \end{dcases*}
\label{eq:k}
\end{equation}
Fig. \ref{Fig:ratio} shows the trap depth geometric factor $\kappa$ as function of $\zeta$. The solid curve in Fig. \ref{Fig:ratio} shows $\kappa$ at a given ion height as a function of $\zeta$ when $b=c$.  The maximum of $\kappa$ can be found at $\zeta \approx 3.68$ for rf electrodes of equal width and for unequal rf electrodes when $c=b/2$ at $\zeta\approx 4.9$ as shown in the dashed curve in Fig. \ref{Fig:ratio}. For the optimised values of $\zeta$, the ion height above the electrodes is given by $h\approx 1.43a$ for the equal width rf electrodes and $h\approx1.27a$ when $c=b/2$. The aim of an optimum trap design is to achieve the maximum trap depth at a given distance above the electrodes. In general, for optimised traps, the maximum trap depth decreases with increasing ion height $h$ and scales approximately as $\sim h^{-2}$. It is important to note that anomalous heating of a trapped ion is proportional to $\sim h^{-4}$ \cite{Turchette,Deslauriers}. The trap depth increases with a decrease in ion-electrode distance, but the heating rate also increases with a decrease in the distance. Therefore, the aim of an optimum trap design is to achieve the maximum trap depth at a given distance above the electrodes.


\section{Optimisation of fast ion separation process}
\label{sec:beta}

In the ion separation process, initially, ions are trapped in a single potential well with secular frequencies $\omega_{x},\omega_{y}$ and $\omega_{z}$, normally ($\omega_{x},\omega_{y}) > \omega_{z}$, where $\omega_{x}$ and $\omega_{y}$ predominantly depend on the pseudopotential provided by the rf electrodes and $\omega_{z}$ depends on the voltages applied to the static potential electrodes. The static voltages can be applied in such a way that a wedge potential can be created between the trapped ions and the single potential well can be pulled apart into two distinct potential wells or a \emph{double well} in the z-direction of the trap geometry. The aim of an effective separation is that the ions remain trapped and acquire minimal kinetic energy during the separation process. Furthermore, we will show that, for adiabatic separation to be possible, the separation process must be performed on the time scale of the minimum secular frequency during the separation.
Following the theoretical work on ion separation by Home and Steane \cite{Home}, the confinement potential near the centre of a trap in the $z$-direction can be analysed using a Taylor expansion as

\begin{equation}
\label{eq:potential}
\Psi(z,t) \approx 2e \alpha(t) z^{2}+2e\beta(t) z^{4}
\end{equation}

\label{separation}
\begin{figure}[!t]
\includegraphics [width=85mm] {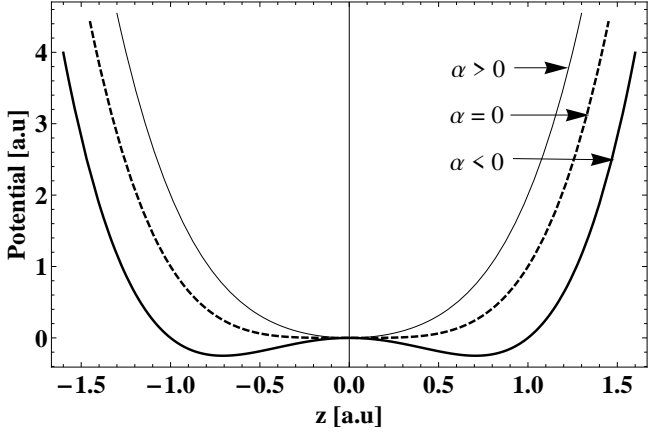}
\caption{Plots of the potential in the $z$-direction when $\alpha > 0$, $\alpha=0$ and $\alpha < 0$. The potential wedge is created when $\alpha < 0$ (solid curve).}
\label{alpha}
\end{figure}

Fig. \ref{alpha} shows the plots of Eq. (\ref{eq:potential}) when $\alpha > 0, \alpha=0 $ and $ \alpha < 0$. A potential wedge can be seen in the middle of the trap when $\alpha$ becomes negative. Secular frequencies, $\omega_z$, in the $z$-direction at each instance can be calculated using the equations derived by Home \emph{et al.} \cite{Home}\\
\begin{equation}
\label{eq:sec}
\omega_{z} \simeq \begin{dcases*}
        \sqrt{\frac{2e\alpha}{m}}  & when $\alpha > 0$ \\ \\
        \sqrt{\frac{3e}{m}}\bigg(\frac{e}{2\pi\epsilon_{o}}\bigg)^{1/5}\beta^{3/10} & when $\alpha \rightarrow 0$\\\\
        \sqrt{\frac{4e|\alpha|}{m}} & when $\alpha < 0$
        \end{dcases*}
\end{equation}

During the separation process, when the voltages $V_i(t)$ on the static potential electrodes are varied in time, the quadruple term $\alpha$ crosses zero and at that point the secular frequency in $z$-direction, $\omega_{z}$, is at its minimum ($\omega_{\text{min}}$). At this point, the ions have a distance of $s\simeq \big(\frac{e}{2 \pi \epsilon_{o}\beta} \big)^{1/5}$ in a single well due to their Coulomb repulsion force \cite{Home}.\\
The value of $\omega_{\text{min}}$ (when $\alpha \rightarrow 0 $) also sets an upper limit on the speed of the separation process. One should therefore aim to maximise $\omega_{\text{min}}$ for a faster separation of ions. When $\alpha\rightarrow0$, $\omega_{\text{min}}$ is only due to the contribution of quartic term $\beta$ in Eq. (\ref{eq:sec}).\\
Therefore summing up the conclusions from Home and Steane \cite{Home}, in order for ion separation to occur, a trap design must provide a negative value of $\alpha$, when appropriate voltages are applied to the trap electrodes. Furthermore, the better trap design is the one which provides higher values for the quartic term $\beta$ which in turn provides a higher value for $\omega_{\text{min}}$ during the separation process allowing for faster speed of the adiabatic shuttling process. From Eq. (\ref{eq:potential}) one can see that, both $\alpha$ and $\beta$ terms depend on applied voltages on the static potential electrodes and their dimensions. As we will discuss in Sec. \ref{sec:OptVolt}, applied voltages on the electrodes are constrained by the power dissipation and the breakdown voltage, therefore, the effective maximum value of $\beta$ needs to be maximised by optimising the trap geometry.\\
\begin{figure}[!t]
\includegraphics [width=80mm] {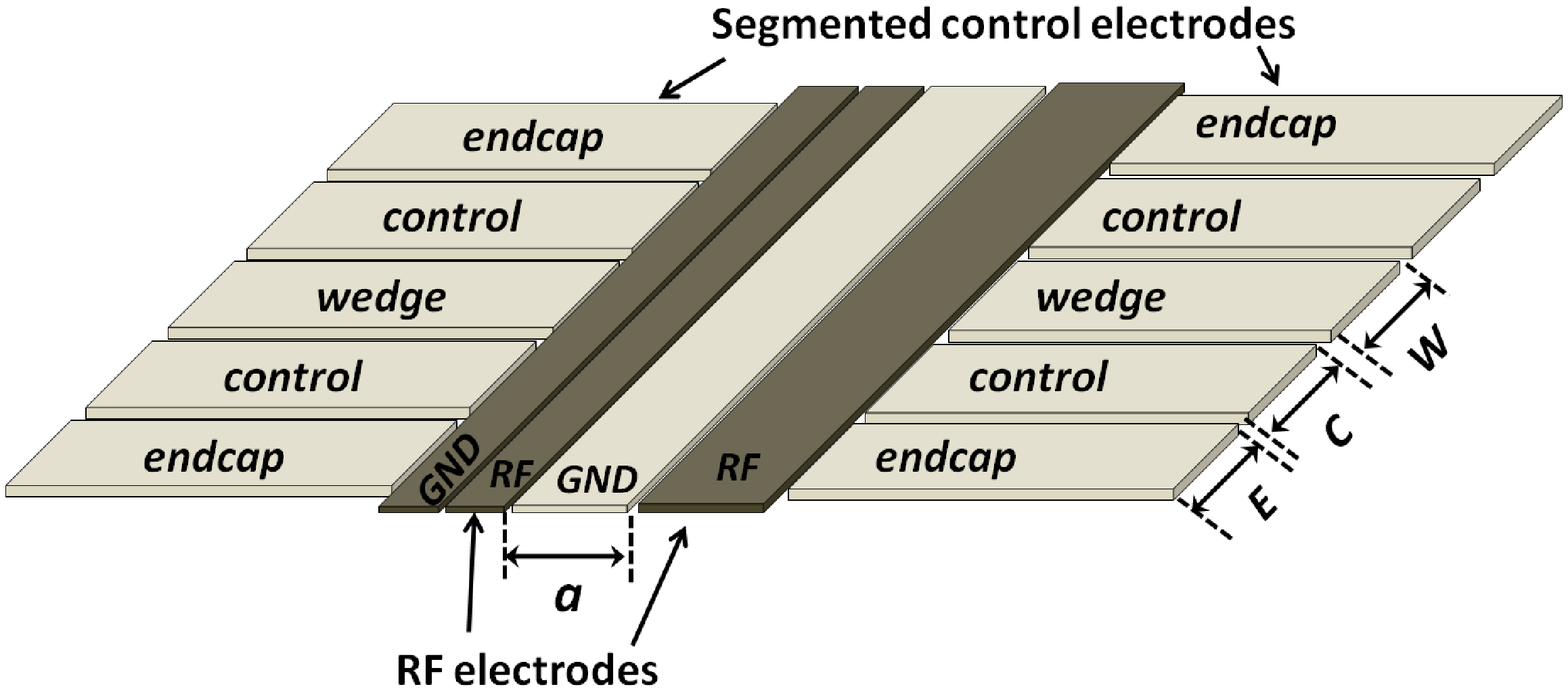} (a)
\includegraphics [width=80mm] {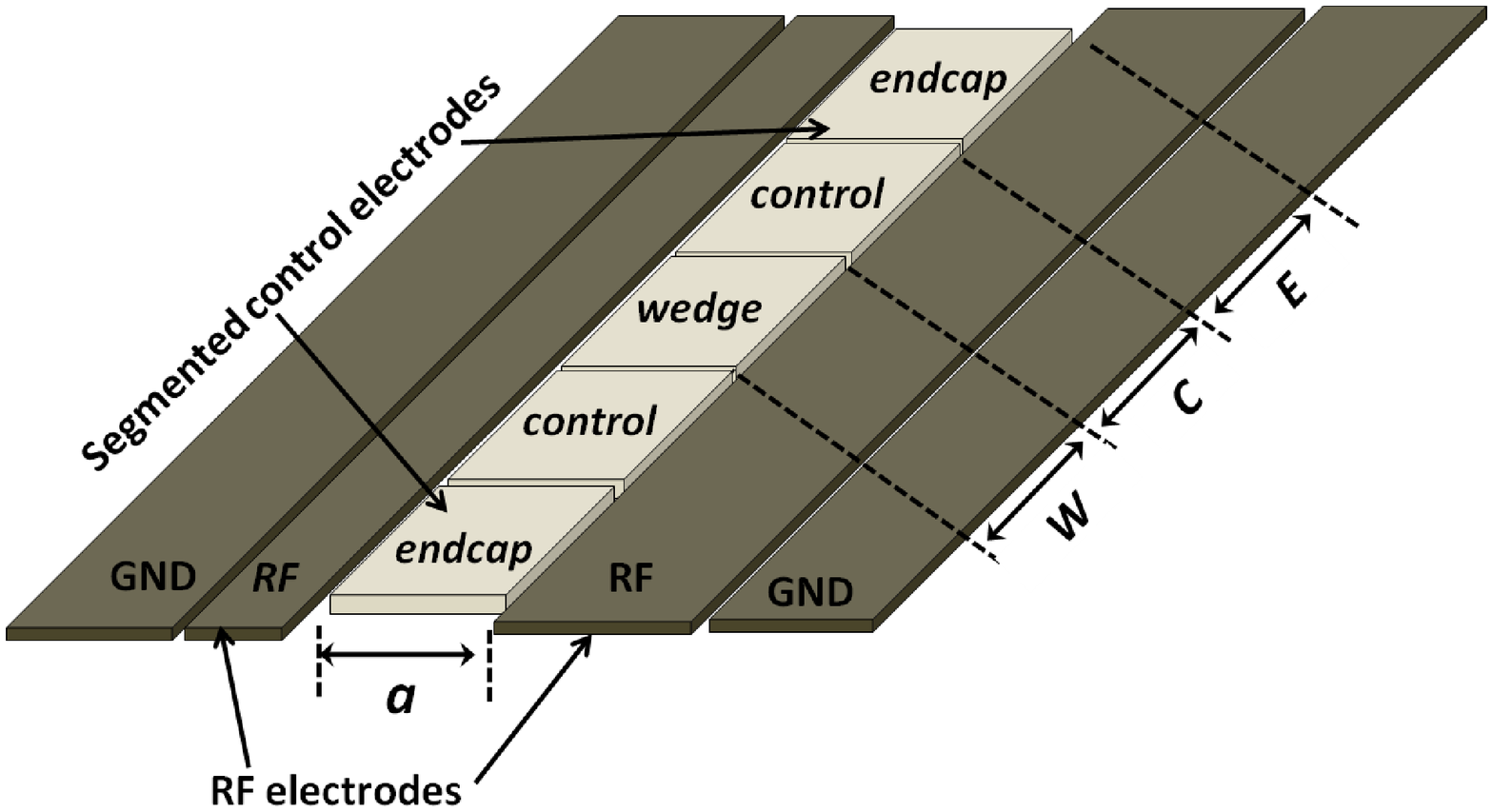} (b)
\caption{Surface trap geometries with (a) outer segmented electrodes and (b) centre segmented electrodes. Unequal rf electrodes are used to provide rotation of the principal axes. }
\label{Fig:Design}
\end{figure}
We investigate two arrangements of static potential electrodes in surface traps for an effective and fast ion separation process. The arrangement of rf and static potential electrodes in the two surface ion trap geometries are shown in Fig. \ref{Fig:Design}. In Fig. \ref{Fig:Design}(a), the outer \emph{rf-ground} electrodes are segmented to provide confinement in the $z$-direction. The ion - static potential electrode distance to the set of static potential electrodes on each side is made equal by inclusion of an additional axial ground electrode. Fig. \ref{Fig:Design}(b) shows an alternative geometry where the central \emph{rf-ground} electrode is segmented to provide axial confinement. The segmented electrodes are labelled as \emph{endcap}, \emph{control} and \emph{wedge}. In both designs, ions can be trapped and separated above the central electrode(s) by adjusting the voltages on the segmented electrodes \cite{Wesenberg,2Chiaverini,Amini}.\\
A surface ion trap lacks one of the reflection symmetries, the symmetry in the direction normal to the surface. Therefore, any applied voltage on the \emph{endcap} electrodes and \emph{wedge} electrodes during separation process can easily alter the position of a trapped ions above the surface \cite{3Pearson} and push the ions out of the \emph{rf node} position. The solution for this problem is to apply a negative voltage on two (in case of the design in Fig. \ref{Fig:Design}(b)) or more (in case of the design in Fig. \ref{Fig:Design}(a)) \emph{control} electrodes symmetrically around the \emph{rf node}. The voltage on the \emph{control} electrodes can be maintained in such a way that the trapped ions always remain in the \emph{rf node} position, during the separation and shuttling processes.

\begin{figure}[t]
\includegraphics [width=85mm] {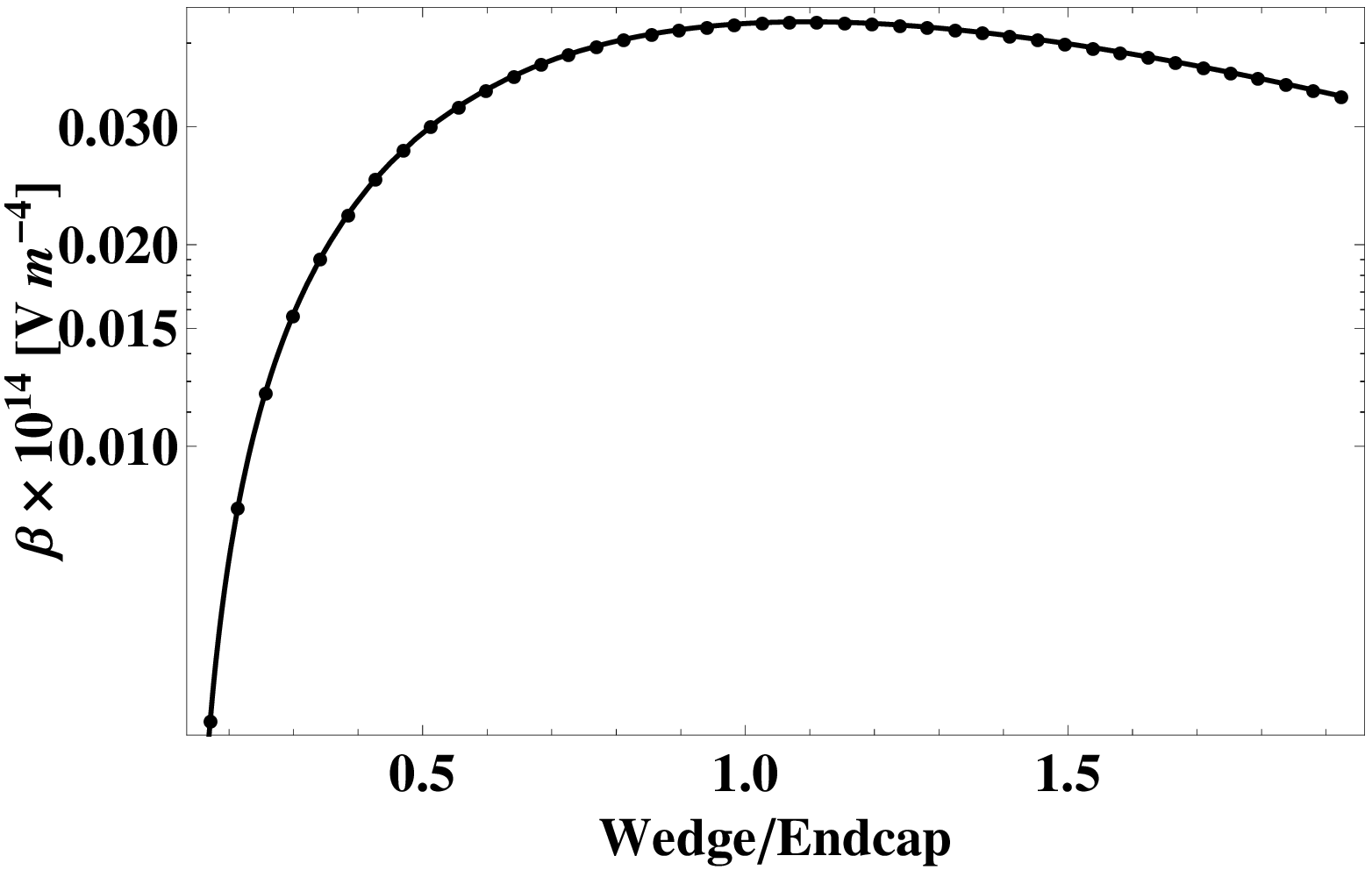}(a)
\includegraphics [width=85mm] {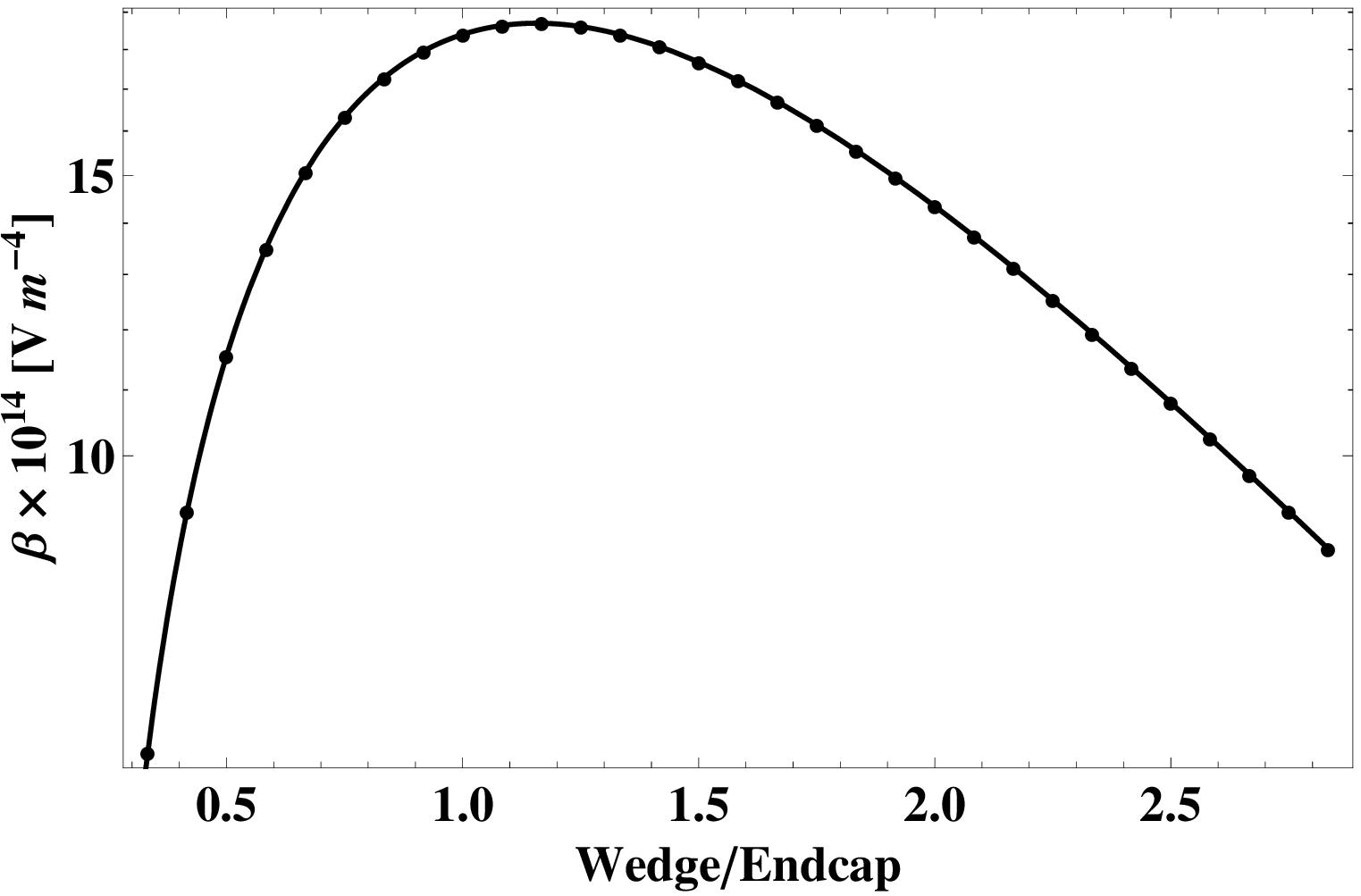}(b)
\caption{ $\beta$ is plotted against the ratio of wedge electrode to endcap electrode width (\emph{W/E})
(a) for the outer segmented electrode design and (b) for the central segmented electrode design.
In both cases an optimal value of \emph{W/E} is approximately 1.1.}
\label{betavsWE}
\end{figure}
\begin{figure}[h]
\includegraphics [width=85mm] {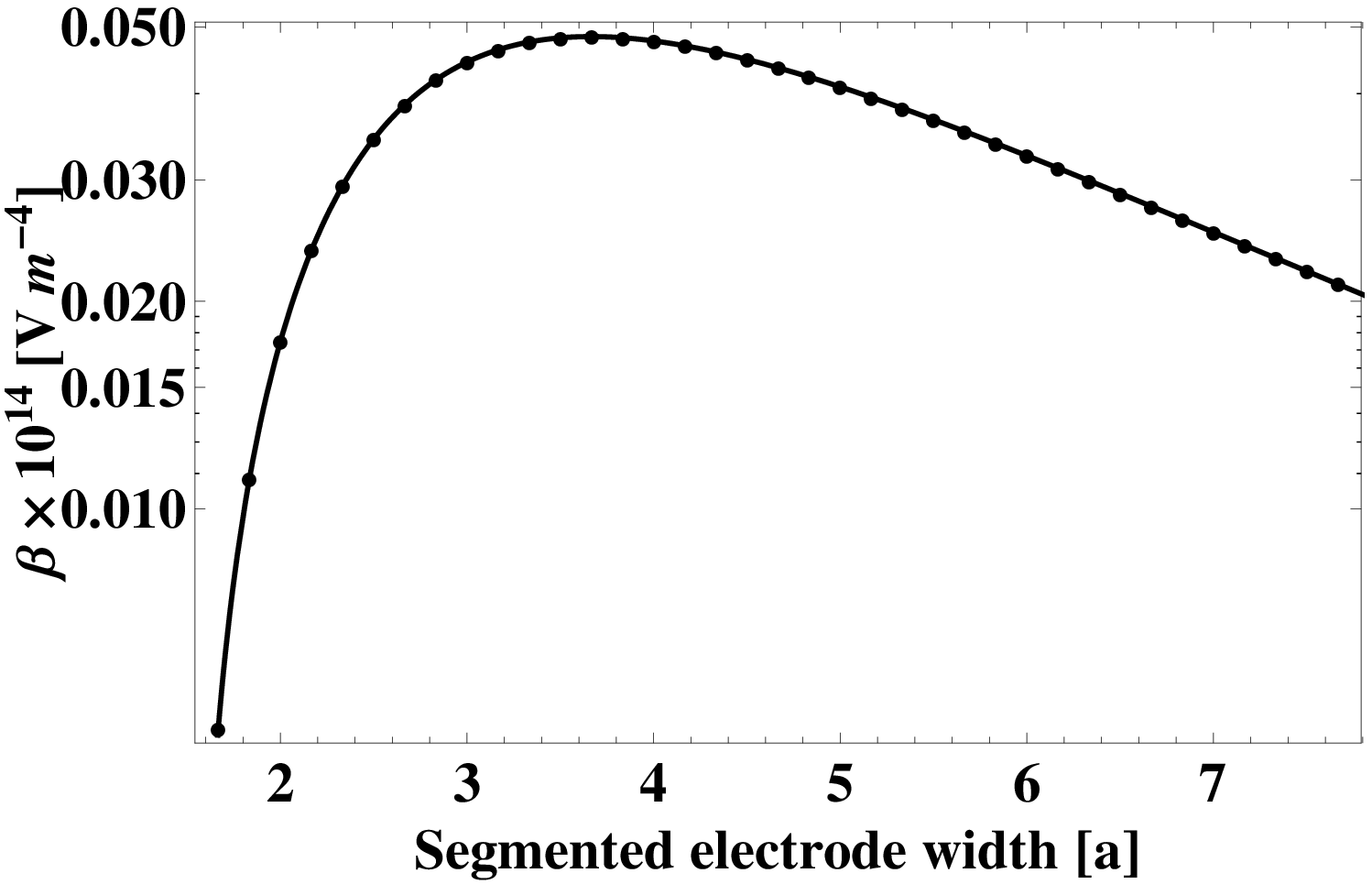}(a)
\includegraphics [width=85mm] {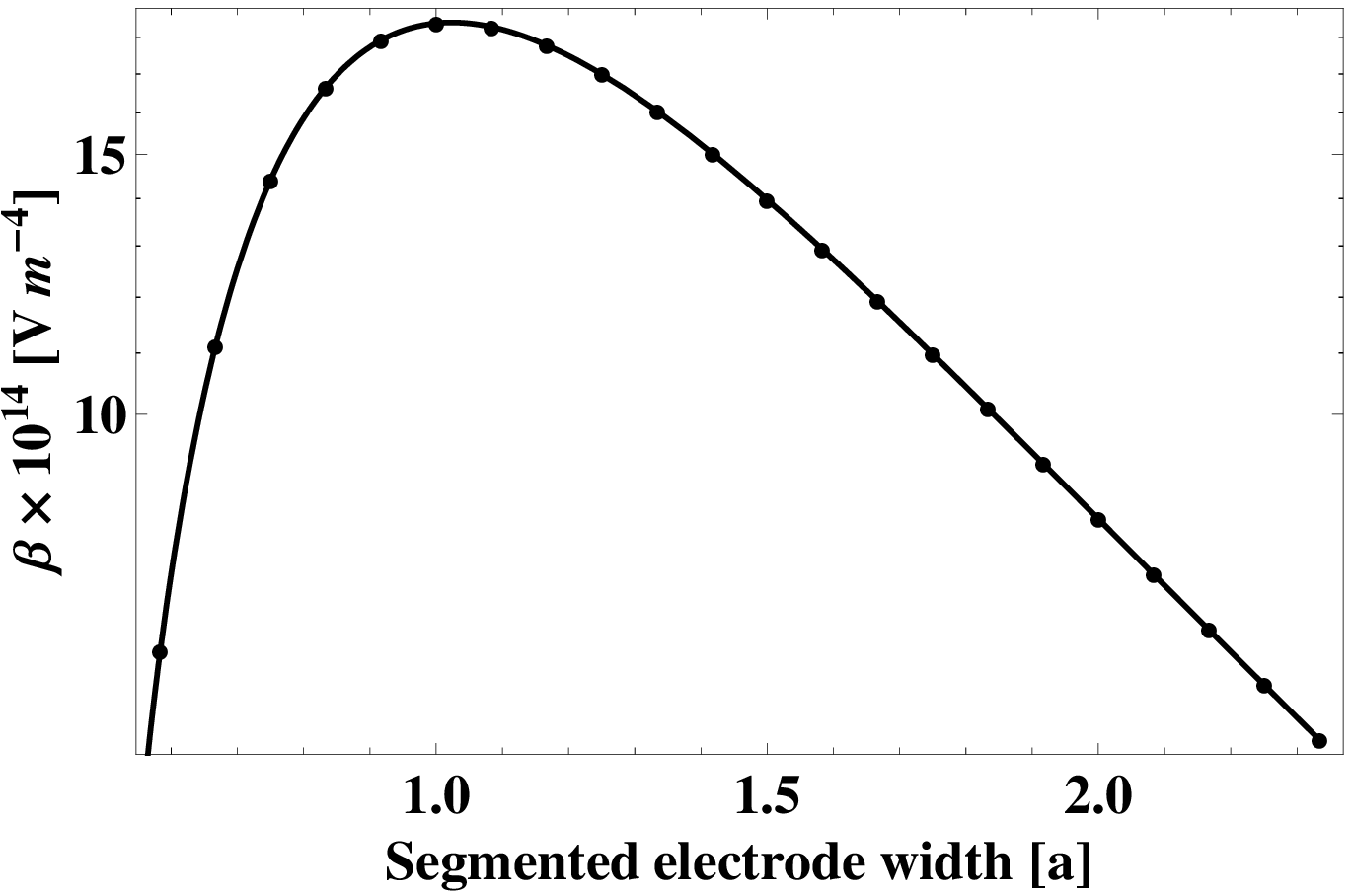}(b)
\caption{ $\beta$ is plotted against electrode widths (\emph{W=C=E}) in units of \emph{a} (a) for the outer segmented electrode design, (b) for the central segmented electrode design.}
\label{betavsa}
\end{figure}
\begin{figure}[t!]
\includegraphics [width=85mm] {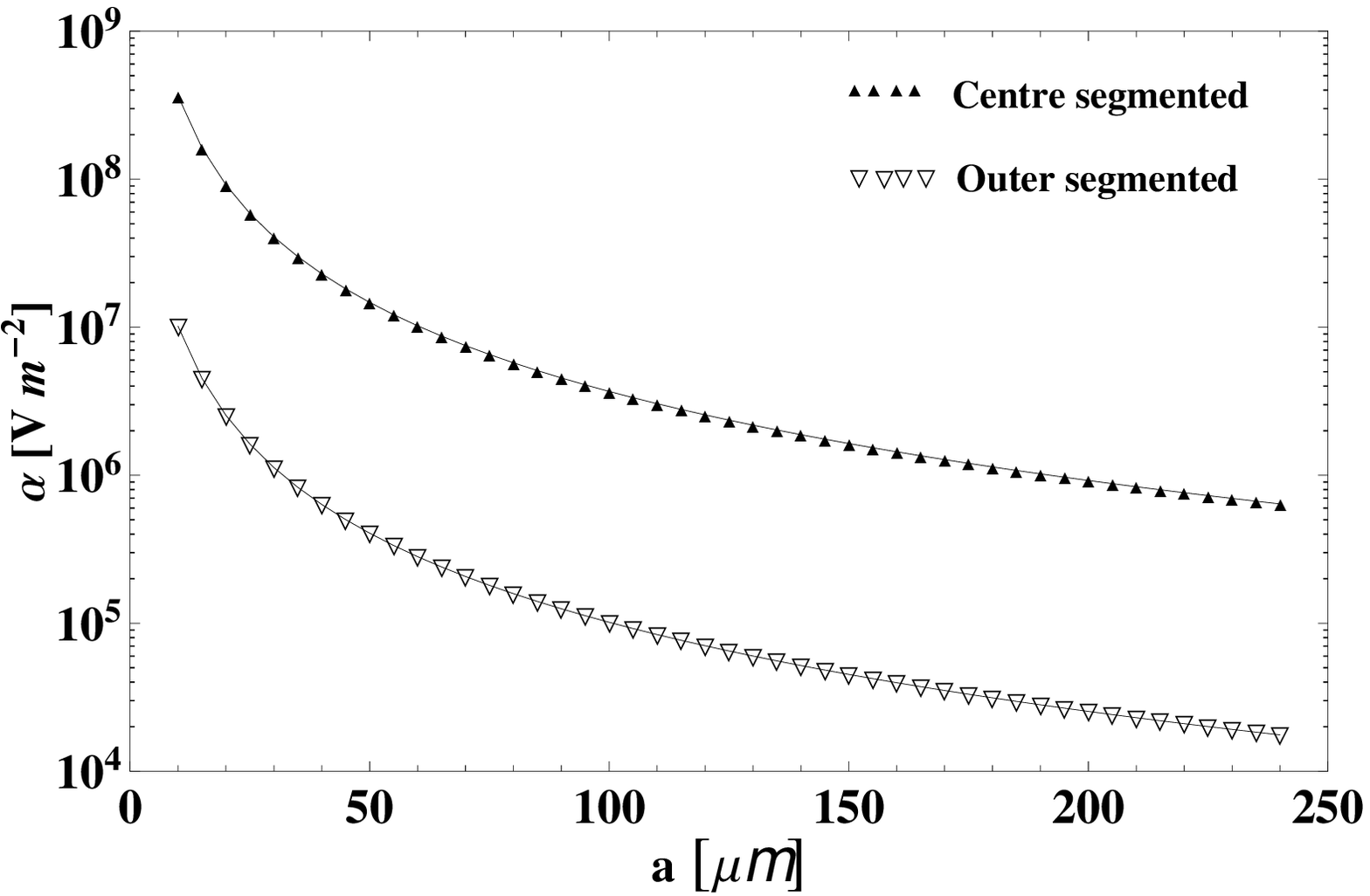}(a)
\includegraphics [width=85mm] {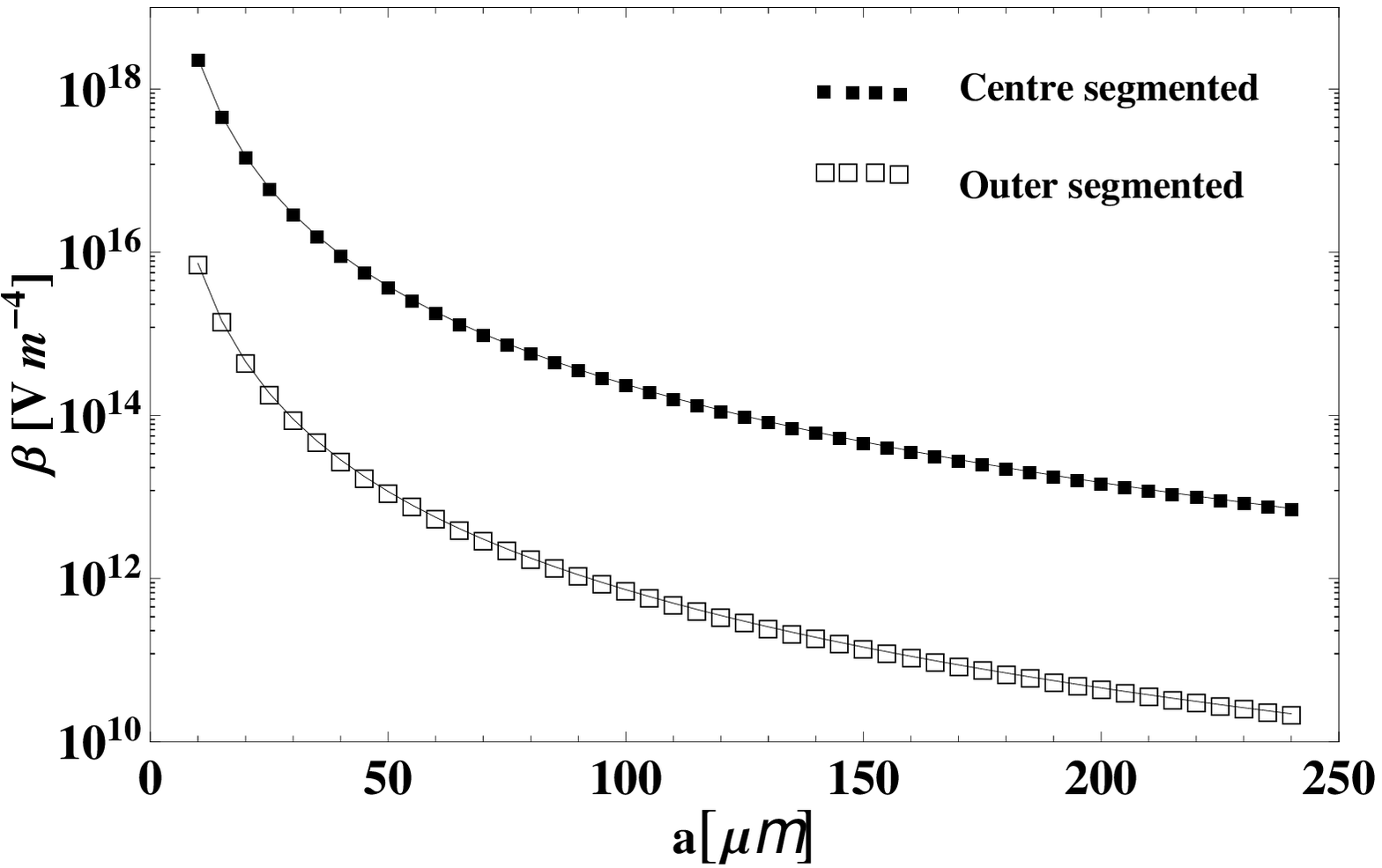}(b)
\caption{ (a) $\alpha$ and (b) $\beta$ for outer and centre segmented geometries are plotted vs \emph{a}. A unit voltage is applied on all trap electrodes. Comparison shows that the $\alpha$ and $\beta$ values are higher in the centrally segmented trap geometry where ion electrode distance is small.}
\label{IED}
\end{figure}
As we know from the previous discussion, an efficient separation process is dependent on the value of $\beta$. Therefore the optimum trap geometry is one which provides maximum values for $\beta$ and allows for negative values for $\alpha$ for certain applied voltages. As the values of the $\alpha$ and $\beta$ parameters are also limited by the voltage constraints of the trap electrodes, in our numerical simulation we keep the applied voltage constant and equal to $1$ V for \emph{endcap} and \emph{wedge} electrodes and $-1$ V for \emph{control} electrodes. This approach allows us to determine the dependency of the $\alpha$ and $\beta$ terms on geometric factors of a surface trap only. The first step is to determine the optimum ratio ($W/E$) of \emph{wedge} to \emph{control} electrode. For this purpose we fix the width of \emph{endcap} and \emph{control} electrodes to equal widths and vary the width of \emph{wedge} electrodes to maximise $\beta$. In Fig. \ref{betavsWE}, $\beta$ is plotted against the ratio of \emph{wedge} to \emph{endcap} electrode widths for both trap designs. In these plots we can see that $\beta$ reaches a maximum when the width of all the static potential electrodes are approximately the same.\\
The widths of static potential electrodes should be chosen in such a way that they provide significant curvature of the potential at the centre of the trap in the \emph{z}-direction and meet the conditions for effective ion separation by providing maximum $\beta$ and negative $\alpha$. In Fig. \ref{betavsa}, $\beta$ is plotted as a function of electrode widths in units of rf electrode separation $a$ for both surface trap designs. We can easily deduce from the plot shown in Fig. \ref{betavsa}(a) that $\beta$ can be maximised when width $W$ of the segmented electrodes is $\approx 3.66 a$ for the design shown in Fig. \ref{Fig:Design}(a). Whilst, in Fig. \ref{betavsa}(b), $\beta$ is maximum when $W\approx a$ for the trap design shown in Fig. \ref{Fig:Design}(b). The relationship between $\beta$ and the segmented electrode width in Fig. \ref{betavsa} for both designs also shows that a relatively larger width of electrodes is required to control the ion motion when the ion-static potential electrode distance is large. The ion - static potential electrode distance, for the outer segmented electrode geometry,  is $\approx \sqrt{[(b+x_o)]^{2}+[h]^2}$, while for the central segmented static potential electrodes, this distance is $\approx h$.\\
Furthermore, in Fig. \ref{IED} we compare both trap geometries for $|\alpha|$ and $\beta$, where both terms are plotted as a function of rf electrode separation $a$ (which dictates the height of the trapping position above the surface). The calculations are made with all other parameters optimised as explained above. By comparing both designs we find that the values for $\alpha$ and $\beta$ are approximately two order of magnitude higher for the design shown in Fig. \ref{Fig:Design}(b). We can also observe a sharp rise in the values of $\alpha$ and $\beta$ when the rf electrode separation decreases below 50 $\mu$m, but the cost for this achievement can be severe because anomalous heating also increases rapidly when the distance between the electrodes and ion decreases.

\section{Ion dynamics problem during separation process}
\label{sec:dynamics}

\subsection{Ion dynamics}
The aim of an optimal separation protocol is to allow the ions to be separated and transported to arbitrary locations within the trap array, whilst ensuring that the motional state of the ion does not change significantely after a shuttling operation. Hucul \emph{et al.} \cite{Hucul} and Reichle \emph{et al.} \cite{Reichle} have identified those constraints that ensure adiabatic transport of ions. Both have highlighted the importance of the inertial forcing of the ions at the beginning and end of a shuttling protocol. Hucul \emph{et al.} showed the benefits of the hyperbolic tangent profile
\begin{equation}\label{eq:Tan}
P_h(t)=\text{tanh}\big[N\frac{t-T}{T}\big],
\end{equation}
and Reichle \emph{et al.} \cite{Reichle} suggested an error function profile
 \begin{equation}\label{eq:Err}
P_e(t)=\text{Erf}\big[n\frac{t-T}{T}\big],
\end{equation}
to shuttle the ions. In both equations (\ref{eq:Tan} and \ref{eq:Err}), $t$ is the instantaneous shuttling time and $T$ is the total shuttling time. Functions with larger $N$ or $n$ produce a more gradual change for values of $t$ close to 0 and $T$, while incorporating larger rates of change for values of $t$ close to $T/2$. If the $N$ and $n$-parameters are selected appropriately, both profiles sufficiently resemble each other. Hyperbolic tangent functions take significantly less computation time than the error function. Therefore we use the hyperbolic tangent time profile to analyse the dynamics of the ion separation processes.\\
An arbitrary time dependent potential can be built using the basis functions for individual electrodes. The force on a charged particle can be calculated using the classical equations of motion \cite{Hucul}
 \begin{equation}
 \label{eq:2ndorder}
\sum_j ^3 m \ddot{\chi_j}+\nabla_j \Psi(\chi_j,t)=0
\end{equation}
where the pseudopotential $\Psi(\chi_j,t)$ is defined in Eq. \ref{eq:iondynamics}. It is also possible to calculate the classical trajectories of ion motion by solving Eq. (\ref{eq:2ndorder}) numerically. High accuracy solutions of Eq. (\ref{eq:2ndorder}) provide the ion dynamics in the $x,\;y$ and $z$ directions as a function of time $t$, which can be used to calculate the kinetic energy gained by the ion. In order to obtain the ion dynamics, we use a package called ``NDSolve" provided in Mathematica-7 to solve these differential equations.

\subsection{Average motional energy}

In a quantum harmonic oscillator of frequency $\omega_o$, the average energy $\big<E\big>$ of level $\big<n\big>$ is given by $\big<E\big>= \hbar \omega \big(\big<n\big>+\frac{1}{2}\big)$. In analogy to a classical harmonic oscillator and assuming the total energy of the ion is only due to its kinetic energy (which is maximum at the bottom of the potential well), the average motional quanta $\big<n\big>_s$ for a trapped ion can be calculated as
\begin{equation}\label{eq:ionenergy4}
\big<n\big>_s= \frac{\frac{1}{2}m v_{t} ^2}{\hbar \omega_{t}}
\end{equation}
where $m$ is the mass of the ion, $v_{t}$ is the maximum velocity and $\omega_{t}$ is the instantaneous secular frequency. The kinetic energy of the ion in the frame of the pseudopotential well is due to its secular motion. By plotting the kinetic energy of the ion versus the shuttling time,
the maximum kinetic energy of the ion at the start and end of the shuttling can be obtained. Hence, the change in the average motional state of the ion is
\begin{equation}\label{eq:ionenergy5}
\big<n\big>_s= \frac{\text{Final}\;K.E_{\text{max}}-\text{Initial}\;K.E_{\text{max}}}{\hbar \omega_{t}},
\end{equation}
where, $K.E_{\text{max}}$ is the maximum kinetic energy of the ion in a potential well.
\subsection{Motional heating caused by anomalous heating}

In a trap design with small ion-electrode distance, significant motional excitation of an ion can be caused by anomalous heating during the shuttling process. It was observed from experimental data that the heating rate $\langle\dot{n}\rangle_{an}$ is related to the ion-electrode distance scaling as $h^{-4}$ and to the secular frequency scaling as $\omega^{-2}$ \cite{Deslauriers}. In order to provide a conservative approximation we can utilise a measurement we have recently carried out \cite{Jim} and the scaling laws as stated above, to provide a realistic estimate of the motional state excitation of the Yb$^+$ due to anomalous heating
\begin{equation}\label{eq:ndt}
\big<\dot{n}\big>_{an} \approx \frac{1.97\pm0.15\times10^{26} \;\mu\text{m}^4\text{Hz}^3}{\omega^2 h^4}
\end{equation}
where $h$ is in units of micrometres and $\omega$ is in units of $s^{-1}$. We note that this expression is only valid for the particular ion trap and ion species it was measured for, however, it provides a reasonable estimate for illustration purposes. We note heating rates can be substantially reduced by operation in a cryogenic environment \cite{Deslauriers,Jaroslaw} as well as optimisation of electrode surfaces and materials.

The motional quanta $\big<n\big>_{an}$ gained from the anomalous heating during the shuttling process can be calculated by integrating $\big<\dot{n}\big>_{an}$ over the shuttling time,
\begin{equation}\label{eq:Mndot}
\big<n\big>_{an}= \int_{t_o} ^{t_f} \big<\dot{n}\big>_{an} dt
\end{equation}
where, $t_o$ and $t_f$ are the start and the end time for a shuttling process. This integral also takes into count the variation of the secular frequency $\omega_z$ during the shuttling process.\\
Therefore, the total number of motional quanta gained during ion transport is given by
\begin{equation}
\big<n\big>=\big<n\big>_{s}+\big<n\big>_{an}
\end{equation}

\subsection{Separation in realistic trap geometries}
\label{sec:OptVolt}

In order to demonstrate the importance of optimal trap geometries for ion separation and recombination, it is useful to analyse the actual dynamics of the process. While the conclusions to be obtained in this section are applicable for geometries featuring a wide range of ion - electrode distances, we carry out actual simulations for a given set of example parameters. Below we motivate the particular choice taken and note that the actual ideal set of parameters should be chosen depending on the particular fabrication process and other considerations such as whether the ion trap is operated in a cryogenic environment, what motional heating rates are acceptable for the particular experiment, what ion species is used and what secular frequencies and trap depths are required.\\
In Sec. \ref{sec:TD} and \ref{sec:beta} we investigated how ion trap electrode dimension ratios can be optimised to maximise the $\kappa$ and $\beta$ parameters which provide for maximum trap depth and secular frequency during separation. In order to carry out simulations of the ion dynamics during separation we need to determine first a realistic set of voltages that can be applied to the chip which will determine the actual values of $\kappa$ and $\beta$ in the shuttling process when using optimised geometries. The trap depth $\Xi$, the rf trap stability factor $q$ \cite{Dehmelt} and the power dissipation $P_d$ in the trap all depend on the applied rf voltage $V_{rf}$, the driving frequency $\Omega_{rf}$ and the given ion mass and the ion-electrode distance \cite{Gosh}. If there is no static voltage offset on the rf electrodes, the trap depth $\Xi$ and the rf trap stability $q$ can be related as
\begin{equation}
\Xi= \frac{V_{rf}}{2\pi^2} \kappa q
\label{eq:td}
\end{equation}
where $q=2eV_{rf}/(m\Omega_{rf}^2 h^2)$ \cite{Gosh} and $\kappa$ is defined in Eq. \ref{eq:k}. Microfabricated ion traps typically are limited in the amount of voltage that can be applied due to voltage breakdown via insulator bulk and surfaces. In order to achieve a deep trap one should therefore choose $q$ as large as possible while still remaining safely inside the region of stability in parameter space. Utilising $q \approx0.7$ seems therefore a reasonable choice. Power dissipation within the ion trap can lead to heating of the trap chip, outgassing from trap material and eventual destruction of the chip. Power dissipation $P_d$ can be estimated as \cite{Marcus}
\begin{equation}
P_d\approx 0.5 V_{rf} ^2\Omega_{rf} ^2 C^2 R
\end{equation}
where, $C$ and $R$ are the trap capacitance and resistance respectively. For typical trap chip configurations a power dissipation of 3 W should not lead to a large temperature change of the ion chip. Considering a typical microfabricated chip with electrodes of thickness $\sim15\;\mu$m made of electroplated gold on a commercially available insulator wafer made of Silicon Oxide (SiO$_2$) (a similar technique is used by Seidelin et al. \cite{Seidelin}), we can estimate typical values for the capacitance and resistance in such traps as $R\approx0.5\;\Omega$ and $C\approx20$ pF, respectively. Setting  $q\approx0.7$ for an $^{171}$Yb$^+$ ion, power dissipation $P_d<3$ W, assuming $\Omega_{rf}\sim2\pi\times 55$ MHz, one could apply $V_{rf}$ of approximately $450 - 500$ V. Microfabricated ion traps typically feature a particular breakdown voltage (eg. largest voltage difference between adjacent rf and static potential electrodes). For this discussion we will assume a maximum voltage difference between adjacent electrodes on the order of 500 V and we will choose voltages applied to the electrodes accordingly.  \\
For the purpose of choosing an illustrative and realistic ion trap geometry we choose an ion-electrode distance sufficiently large to feature reasonably low motional excitation due to anomalous heating while allowing for reasonable trap depth and secular frequencies when realistic voltages are applied to the trap. If we choose an ion height of $\approx85\;\mu$m, a trap depth $\Xi\approx0.32$ eV and radial secular frequencies ($\omega_x$ and $\omega_y$) of up to $\sim4.2$ MHz for an $^{171}$Yb$^+$ ion can be achieved by using the parameters mentioned above. Having chosen the height of the ion above the surface, the optimisation considerations in Sec. \ref{sec:TD} and \ref{sec:beta} uniquely determine all other electrode dimensions. The optimum rf electrode widths assuming unequal rf electrodes widths ($b=c/2$), chosen for principal axis rotation, are therefore $b\approx300$ $\mu$m and $c\approx150$ $\mu$m separated by a ground electrode of width 50 $\mu$m (i.e. $a=60\;\mu$m including the gaps of $5\;\mu$m). The optimum width of the control electrodes for the ion separation process are then $W=C=E\approx220$ $\mu$m for the design shown in \ref{Fig:Design}(a) and $W=C=E\approx60$ $\mu$m for the design shown in \ref{Fig:Design}(b). \\
Using these parameters we solve the equations of motion to calculate the ion dynamics and resulting overall motional excitation. While the actual results correspond to the particular trap parameters stated above, the conclusions obtained are applicable for all surface ion trap arrays. For simplification we will refer to the centrally segmented design (Fig. \ref{Fig:Design}(a)) and the outer segmented design (Fig. \ref{Fig:Design}(b)) with constraints and dimensions explained above as Centre Segmented Trap and Outer Segmented Trap, respectively.
\\

\subsection{Ion separation in the outer segmented trap }

\begin{figure}
\center
\includegraphics [width=85mm] {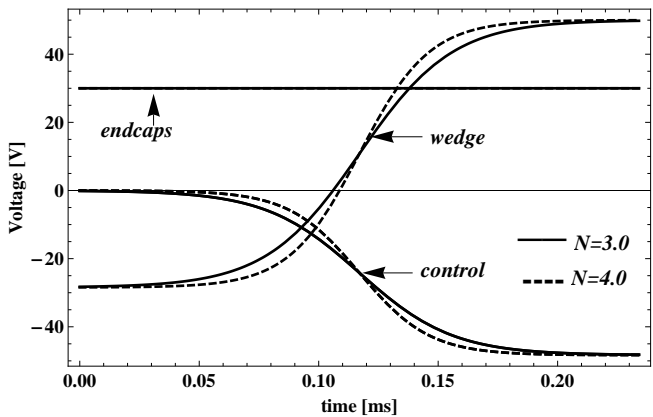}
\caption{Voltage variation as a function of time for the \emph{control} and \emph{wedge} electrodes to produce the ion separation process in the outer segmented trap.}
\label{fig:MSepVO}
\end{figure}
\begin{figure}[!h]
\center
\includegraphics [width=85mm] {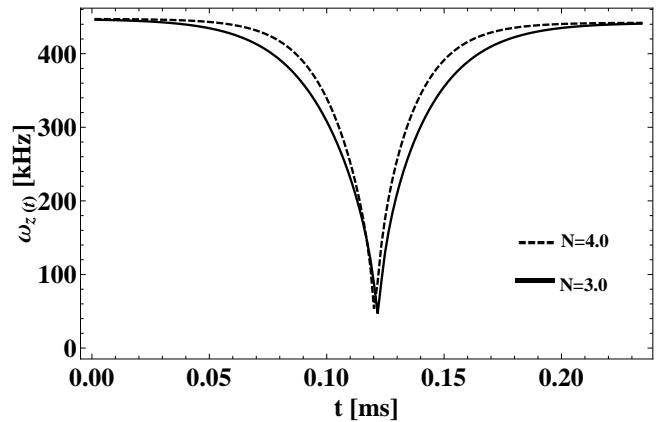}
\caption{Variation of the secular frequency $\omega_z$ during the ion separation process in the outer segmented trap using the hyperbolic tangent time profile with $N=3.0$ (dashed) and $N=4.0$ (solid). The secular frequency at the start and the end of the process is $\omega_{z}/2\pi\approx 500$ kHz. The lowest secular frequency experienced by the ions during the separation process is $\omega_{min}/2\pi \approx 42$ kHz.}
\label{fig:MSepSecO}
\end{figure}
\begin{figure}[!b]
\center
\includegraphics[width=85mm]{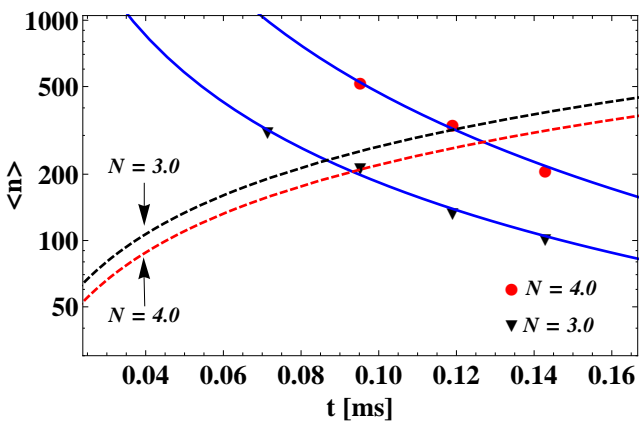}
\caption{Gain in the average motional quanta $\big<n\big>$ by the ion in the External Segmented Trap after the separation process as a function of shuttling time. Hyperbolic tangent time profiles using $N=4.0$ and $N=3.0$ are used to change the voltage on the control electrodes. The solid lines represent the best fit to the average motional quanta $\big<n\big>_s$ resulting from the shuttling process and the dashed lines shows the gain of $\big<n\big>_{an}$ from motional heating in the trap. The crossing points set lower limits for the total gain in $\big<n\big>$ during the shuttling process.}
\label{fig:MSepNO}
\end{figure}

First we discuss the dynamics in the outer segmented trap (Fig. \ref{Fig:Design}(a)). We confine $^{171}$Yb$^+$ ions using $V_{rf}\approx450$ resulting in radial secular frequencies $\omega_x\approx\omega_y\approx4$ MHz. Axial confinement along the $z$-direction is obtained by applying the static voltage of approximately +30 V on the \emph{endcap} electrodes and approximately -34 V on the \emph{wedge} electrodes resulting in a secular frequency in the $z$-direction of $\omega_z\approx440$ kHz. Using these static voltages maximises the axial secular frequency while still retaining sufficient trap depth. The ions are separated by changing the voltage on the \emph{wedge} electrodes to approximately +50 V and the voltage on the \emph{control} electrodes to approximately -48 V monotonically. At the end of the separation process, the ions are located in two distinct potential wells approximately $2 \times W$ apart. The static voltages are chosen to achieve maximum secular frequencies during the separation process and to retain sufficient trap depth (at least 0.2 eV) during the transport.\\
We use the hyperbolic tangent time profile for changing the voltage on the static potential electrodes in the ion separation process. To illustrate the functionality of the $N$-parameter in the separation process, we use $N=3.0$ and $N=4.0$.  Fig. \ref{fig:MSepSecO} shows the voltages applied to the electrodes as a function of time. \\
As discussed in Sec. \ref{sec:beta}, the secular frequency in the $z$-direction varies during the separation process. The variation of the secular frequency $\omega_z$ for an $^{171}$Yb$^+$ ion during the separation process followed by hyperbolic tangent time profile with $N=3.0$ and $N=4.0$ is plotted in Fig. \ref{fig:MSepSecO}. We can see that the secular frequency reaches a minimum $\omega_{min}\approx 2\pi\times42$ kHz when the double well is about to appear. As shown in Fig. \ref{fig:MSepSecO}, the secular frequency of the ions varies rapidly and at one point is at its lowest. Therefore, to reduce the energy gain during the separation process, the ions should be separated slower than the time scale of the minimum secular frequency $2 \pi /\omega_{\text{min}}$.
The gain in motional quanta $\langle n\rangle$ is plotted in Fig. \ref{fig:MSepNO} as a function of shuttling time for hyperbolic tangent with $N=4.0$ and $N=3.0$. The solid curves represent the motional quanta gain due to the shuttling process $\big<n\big>_{s}$, whilst the dashed curves show the motional quanta gain caused by anomalous heating $\big<n\big>_{an}$. Note we use a particular heating scaling law (eq. \ref{eq:ndt}) which is only valid for a particular ion trap and ytterbium ions. We only use this law for illustration purposes, the use of other ions species and trap materials will result in different absolute values of motional excitation even though the observed trends will remain the same. The crossing points between these two curves provide a reasonable idea of the minimum motional excitation that can be achieved. While longer shuttling times will reduce the motional excitation that results from the actual shuttling process, they will increase motional excitation due to anomalous heating. Therefore finding the crossing point between the two curves provides for the optimal shuttling time scale. We stress that the actual minimum achievable excitation in a particular shuttling process is dependent on the actual motional heating rate, the figures here only serve to illustrate the principle.\\
The gain of the motional quanta $\big<n\big>_{an}$ also depends on the $N$-parameter of the hyperbolic tangent profile. This can be explained by the profile of the variation of the axial secular frequency during the separation process with $N=3.0$ (solid curve) and $N=4.0$ (dashed curve) shown in Fig. \ref{fig:MSepSecO}. The graph shows that the ions spend a relatively long time in the lower frequency region during the separation process when the value of the $N$-parameters is smaller therefore being subject to more motional excitation via anomalous heating.  Fig. \ref{fig:MSepNO} shows that approximately the same number of quanta is added at the crossing points for the separation profiles with $N=3.0$ and $N=4.0$, but the duration of the separation is smaller in case of $N=3.0$. Optimising for the best value of $N$ allows for small gains in achievable lowest motional excitation and shuttling speed.

\subsection{Ion separation in the centre segmented trap }
\begin{figure}
\center
\includegraphics [width=85mm] {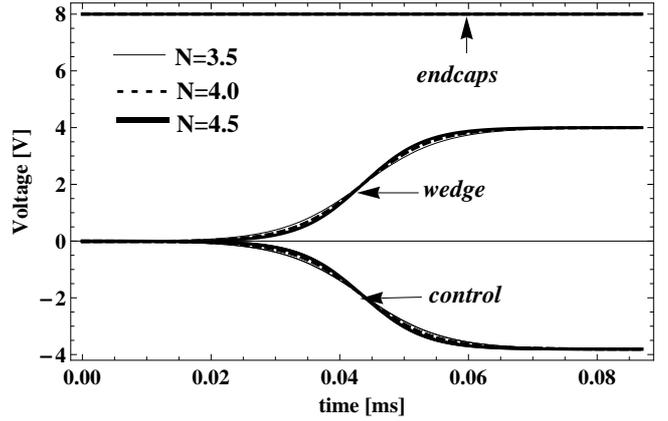}
\caption{Voltage variation as a function of time for the \emph{control} and \emph{wedge} electrodes to produce the ion separation process for the Centre Segmented Trap.}
\label{fig:MSepV1}
\end{figure}
\begin{figure}[!b]
\center
\includegraphics [width=85mm] {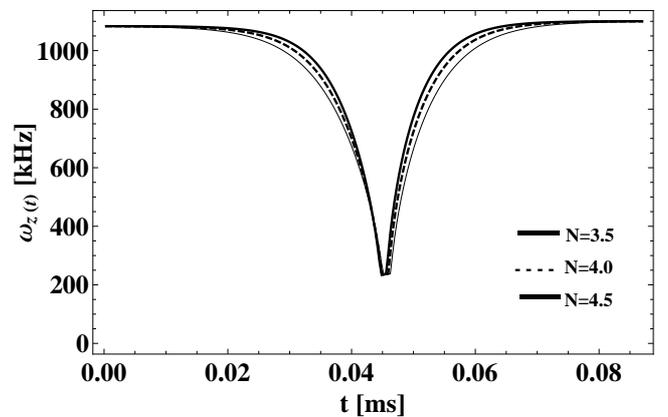}
\caption{Variation of the secular frequency $\omega_z$, during the ion separation process in the Centre Segmented Trap using a hyperbolic tangent time profile of $N=3.5$, $N=4.0$ and $N=4.5$. The secular frequency is $\omega_{z}/2\pi\approx 1.15$ MHz at the start and at the end of the process. The lowest secular frequency attained by the ions during the separation process is $\omega_{min}/2\pi \approx 230$ kHz.}
\label{fig:MSepVC}
\end{figure}

\begin{figure}
\center
\includegraphics[width=80mm]{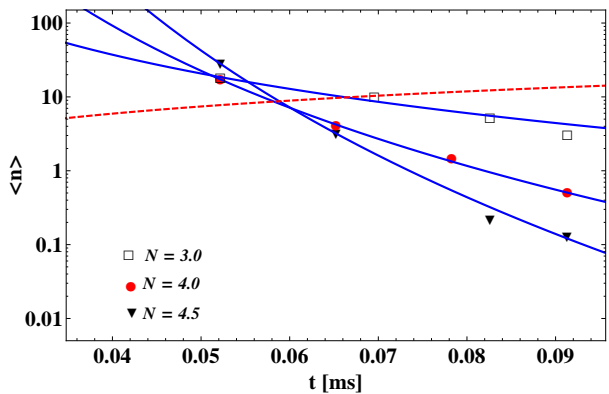}
\caption{Gain in the average motional quanta $\big<n\big>$ by the ion in the Centre Segmented Trap after the separation process as a function of shuttling time. Hyperbolic tangent time profiles using $N=4.0$ and $N=3.0$ are used to change the voltage on the control electrodes. The solid lines represent the best fit to the average motional quanta $\big<n\big>_s$ resulting from the shuttling process and the dashed lines shows the gain of $\big<n\big>_{an}$ from motional heating in the trap. The crossing points set lower limits for the total gain in $\big<n\big>$ during the shuttling process.}
\label{fig:MSepNC}
\end{figure}

Next we discuss the centre segmented trap illustrated in Fig. \ref{Fig:Design}(b). Ytterbium ions are initially stored in a single potential well applying $V_{rf}\approx 500$ V and a static voltage of approximately 8 V on the \emph{endcap} electrodes. The static voltages are chosen to confine the ions with approximately maximum secular frequency while not making the overall potential anti-trapping and retaining at least 0.2 eV overall trap depth. With these voltages applied, the maximum secular frequency in axial direction is $\omega_z\approx1$ MHz and the radial secular frequencies are $\omega_x\approx\omega_y\approx4.3$ MHz. We separate the ions by adjusting the voltage on the \emph{wedge} electrode to approximately 4 V and the voltage on the \emph{control} electrodes to approximately -3.8 V. In order to optimise the $N$-parameter, we carry out simulations using the hyperbolic tangent time profile with $N=(3.0,\;4.0,\;4.5)$. Similarly as in the case for the external segmented geometry, the secular frequency in z-direction varies during the separation process as shown in Fig. \ref{fig:MSepVC}. The lowest secular frequency $\omega_{min}$ for an Yb$^+$ ion during the separation process is approximately 230 kHz. Fig. \ref{fig:MSepV1} shows the voltages applied to the electrodes as a function of time.\\
The average motional quanta $\langle n\rangle$ gained by the ion after the separation process are plotted as a function of total shuttling time in Fig. \ref{fig:MSepNC}. The solid curves represent the motional quanta gain due to the shuttling process $\big<n\big>_{s}$, whilst the dashed curves show the motional quanta gain caused by anomalous heating $\big<n\big>_{an}$. The crossing points of $\big<n\big>_{s}$ (solid lines) and the $\big<n\big>_{an}$ (dashed line) set the lower limit of the total average motional quanta $\big<n\big>$ gained by the ion during the separation process. There are only minor differences for the different $N$ parameters.

\subsection{Comparison}
By analysing the actual dynamics of the ion separation process using realistic examples we can learn a lot about optimal separation. While one may assume optimal ion trap geometries are only useful to provide for faster adiabatic separation processes, our results show that optimal geometries may in fact be a prerequisite for adiabatic separation due to the existence of anomalous heating caused by fluctuating charges on the trap electrodes. Furthermore, we show that the minimum gain of total motional quanta $\big<n\big>$ during the separation process is much lower in the centre segmented electrode geometry due to the higher achievable values of the axial secular frequency $\omega_{z}$ during the separation process. In fact, the centre segmented geometry allows for much faster separation with overall motional excitation still remaining very small. It is also possible to achieve higher secular frequencies in the centre segmented geometry while still retaining sufficient overall trap depth. \\
These results also demonstrate the importance of optimisation of electrode dimensions as derived in Sec. \ref{sec:TD} and \ref{sec:beta}. A geometry with optimised trap depth provides the capability for applying larger static voltages (before the overall potential becomes anti-trapping), due to the deeper trap depth which in turn provides for larger axial secular frequencies and faster ion separation. Particularly in the case of the outer segmented trap geometry, it is important to use optimised electrode dimensions in order to partially compensate for the in-principle limitations caused by anomalous heating in order to establish at least near adiabatic operation.  We stress that the actual values of estimated total motional excitation only serve illustrative purposes and are expected to significantly vary when using different ion species and ion traps.
\section{Conclusion}
We have demonstrated that effective and fast ion separation in scalable surface ion traps at maximum trap depth can be achieved by optimising sizes and arrangement of the ion trap electrodes within a surface ion trap array. We have calculated the optimum ratio of the widths of rf electrodes over their separation for the maximum trap depth at a given ion height. The trap parameters $\alpha$ and $\beta$ which characterise the secular frequencies during the separation process can be maximised by optimisation of the electrode dimensions. We have solved the equations of motion for the dynamics of ion separation and illustrated the importance of optimised electrode configurations. A separation process performed with higher secular frequencies adds less amount of energy to the ions. We have shown that centrally segmented ion trap geometries are superior in their performance compared to outer segmented geometries. Centrally segmented geometries allow for significantly smaller overall motional excitation and also provide for much faster adiabatic separation processes. In fact, depending on the actual experimental conditions, they may even be a prerequisite to accomplish adiabatic separation. Due to the much simpler fabrication of outer segmented geometries, these may nevertheless be a geometry of choice. In that case our article illustrates both the importance and design such a geometry with optimal trap dimensions.  \\
Ion trap arrays are of significant importance for the implementation of scalable quantum technology with trapped ions. Ion separation within such arrays may likely play a critical role and our paper shows how this process can be accomplished optimally.

\section*{ACKNOWLEDGMENTS}
This work was supported by the UK Engineering and Physical Sciences Research
Council (EP/E011136/1, EP/G007276/1), the European Commission's Sixth
Framework Marie Curie International Reintegration Programme
(MIRG-CT-2007-046432), the Nuffield Foundation and the University of Sussex.



\end{document}